# BACK-PROPAGATION OF ACCURACY

*Masha Yu. Senashova*
Krasnoyarsk State University

*Alexander N. Gorban*
Krasnoyarsk Computing Center RAS
660036, Krasnoyarsk-36, Computing Center RAS, Russian Federation
E-mail: gorban@cc.krascience.rssi.ru

*Donald C. Wunsch II*
Applied Computational Intelligence Laboratory
Department of Electrical Engineering, Texas Tech University
Lubbock, TX 79409-3102
E-mail: Dwunsch@aol.com

In this paper we solve the problem: how to determine maximal allowable errors, possible for signals and parameters of each element of a network proceeding from the condition that the vector of output signals of the network should be calculated with given accuracy? "Back-propagation of accuracy" is developed to solve this problem.

In the present time there are various technical realizations of neural networks, including neural network simulations which are computer models of neural networks. Neural network simulations are flexible means for training of networks and working with them. It is possible to carry out various operations with neural network simulations: to train, to determine the most and the least significant connections, to contrast, i.e. to remove the least significant connections etc. We will consider trained neural network simulation, i.e. for which the values of synapse weights are calculated. Neural network simulation working on digital computer allows to calculate synapse weights with high accuracy, what is difficult to obtain with other technical realizations of network because of their limited accuracy. Therefore there is the problem of binarization of synapse weights, that is reduction of synapse weights to some set of particular values (for example, 0 or 1 – absence or presence connection between neurons, -1 or 1 etc.). It is attained by change of the architecture of network – by addition of new synapses, new summators, by creation of cascades of summators etc. On binarization of a network, it is technically difficult to obtain the result with the same degree of accuracy as the result of neural network simulation, because sometimes for exact realization it is required about thousands, tens of thousands or more connections instead of one initial. Therefore one should beforehand determine some accuracy of the result of network operation after binarization, that is to choose an interval of variations of the values of the vector of output signals of the network.

The effect of weight and input signal errors to functioning of neural networks was study in [3,4,5].

The idea of the method has arisen when solving the problem of binarization of neural network. But back-propagation of accuracy is interesting not only in the application to a problem of binarization. It can be applied for solving a number of other problems. For example, having calculated allowable errors for the whole network, one can find out, in which limits it is possible to vary input data and signals on any part of networks in order that the vector of output signals would change within the given limits.

In addition we assume that the network has layered structure. A network of layered structure consists a number of layers of standard neurons, interconnected by synapses with the weights calculated in training. The signals are transferred only in one direction, from the previous layer to the following. Standard neuron we understand as a set of elements, which are adaptive summator, nonlinear transformer and branch point (Fig.1.). The branch point is an element sending output signal of the nonlinear transformer to inputs of several standard neurons of the following layer. Description of structures of such type see in [1,2].

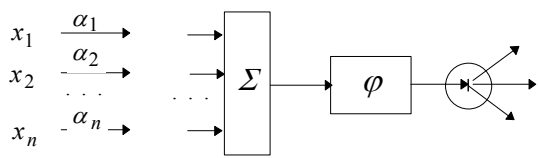

*Fig.1. Standard neuron.*
*Σ — adaptiv summator, φ — nonlinear transformer and branch point.*

Let for each components $y_i$ of the vector of output signals of network be given the allowable error of calculation $\delta_i$.

Since we deal with the networks of layered structure consisting of layers of standard neurons, the output signals of one layer are the input signals of other layer. In turn, the output signal of one element (for example, summator) is the input signal of other element (for example, nonlinear transformer) inside standard neuron. Thus, starting from the output signals of the network it is possible to find out from which element of the network the given element has received the signal. Since the allowable errors of calculations for output signals of the network are given, it is possible to calculate allowable errors for each element, passing from element to element in reverse direction.

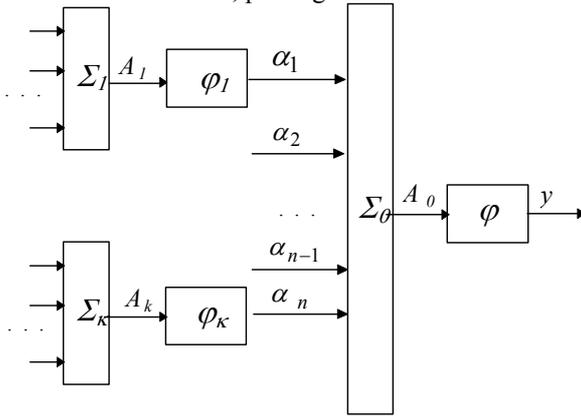

*Fig.2. $\Sigma_i$ – summators, $A_i$ – summator output signals, $\varphi$, $\varphi_i$ – nonlinear transformers, $\alpha_i$ – synapse weights.*

Typical part of network is standard neuron. Any neural network consists of a number of standard neurons. Therefore it is enough to find out how are calculated allowable errors for elements of standard neuron. Then we will have possibility to calculate allowable errors for any part of network.

Let consider the problem of distribution of accuracy for the following typical part of a network, consisting of the summator $\Sigma_0$ and nonlinear transformer, the result of work of which is a output signal $y$ (here and further $y$ denotes component of the vector of output signals of network), and also of summators $\Sigma_i$ and nonlinear transformer, the output signals of which are the input signals of the summator $\Sigma_0$ (Fig.2). Actually we consider two last layers of neural network consisting of standard neurons.

Let directly consider the method. We have to calculate allowable errors for each element of this part of the network, recognizing that the output signal $y$ changes not more than at $\delta$.

Since the last element of standard neuron is the branch point, let begin consideration of back-propagation of accuracy just from it.

The errors $\varepsilon_i$ coming to the branch point during back-propagation may be unequal. Choosing as an error the smallest value, we can only increase the accuracy of calculations. Thus, it is necessary to choose $\varepsilon = \min\{\varepsilon_i\}_{i=1}^{n}$ as the allowable error, which should leave the branch point for further calculation of errors. At the considered part of network (Fig.2) standard neuron of the last layer does not include branch point, since the output signal of the nonlinear transformer is simultaneously the output signal of the network.

The following element of standard neuron is the nonlinear transformer with input signal $A_0$, activation function $\varphi$ and output signal $y = \varphi(A_0)$. Let calculate the maximal error $\varepsilon$ of input signal of the nonlinear transformer, that is find an interval $[A_0-\varepsilon, A_0+\varepsilon]$ such that for any $x \in [A_0-\varepsilon, A_0+\varepsilon]$ $\tilde{y} = \varphi(x)$ differs from $y = \varphi(A_0)$ not more than by $\delta$: $|\tilde{y} - y| \leq \delta$.

Owing to continuity and differentiability of activation function of nonlinear transformer it is obvious that $\varepsilon \leq \delta / \max|\varphi'(x)|$, where $x \in [\varphi^{-1}(y-\delta), \varphi^{-1}(y+\delta)]$.

According to traditional way, let estimate allowable error in linear approximation: $\varphi(A_0 \pm \varepsilon) \approx \varphi(A_0) \pm \varphi'(A_0) \cdot \varepsilon$. We can select $\varepsilon$ as follows: $\varepsilon \leq \delta / |\varphi'(A_0)|$. In this case the formula for calculation of allowable errors is more simple but less exact.

We have obtained an error, allowable for input signal of nonlinear transformer, which simultaneously is allowable error for the result of work of the summator $A_0$. We can similarly calculate the error of input signal of nonlinear transformer of any standard neuron if the error of its output signal is known.

Come to the following element of standard neuron – the adaptive summator with $k$ synapses being its inputs. The adaptive summator is the summator, in which the input signals $x_j$ are summarized with weights $\alpha_j$.

Each input $x_i$ of the summator $\Sigma_0$ also has some error $\varepsilon_i$. For uniform distribution of error we assume that all inputs of the summator have equal accuracies ($\varepsilon_1 = \varepsilon_i$, $i=2,...,k$).

Let $A_0 = \sum_{i=1}^{k} \alpha_i \cdot x_i$ be the result of work of initial summator. Then $\{A_0'\}$ is the results of work of the summator, when the vector of input signals of the summator comes through the vertices of $k$-dimensional cube with center in the point $(x_1, x_2, ..., x_k)$ and edge of length $2 \cdot \varepsilon_1$:

$$\{A_0'\} = \sum_{i=1}^{k} \alpha_i \cdot (x_i \pm \varepsilon_i) = \sum_{i=1}^{k} \alpha_i \cdot x_i + \sum_{i=1}^{k} \alpha_i \cdot (\pm \varepsilon_i) = A_0 + \sum_{i=1}^{k} \alpha_i \cdot \varepsilon_i \cdot z_i ,$$

where $z_i = \{-1, 1\}$. In order all the set $\{A_0'\}$ to belong to the interval $[A_0 - \varepsilon, A_0 + \varepsilon]$ it is necessary that $\max \left| \sum_{i=1}^{k} \alpha_i \cdot \varepsilon_i \cdot z_i \right| = \sum_{i=1}^{k} |\alpha_i| \cdot \varepsilon_i \leq \varepsilon$, where the maximum is taken through all $z_i$. From this inequality and the above assumption about $\varepsilon_i$ we obtain the required estimation for $\varepsilon_i$: $\varepsilon_i \leq \varepsilon / \sum_{i=1}^{k} |\alpha_i|$.

The distribution of allowable errors through inputs of the summator such that the error of each input is calculated using the formula $\varepsilon_i \leq \varepsilon / \sum_{i=1}^{k} |\alpha_i|$, is called uniform.

In addition to uniform distribution of errors through inputs of the adaptive summator it is possible also to use proportional and priority distribution.

Under proportional distribution of errors the summator allowable error is divided at first by the number of inputs, and then for each input it is divided by corresponding synapse weight. That is, under proportional distribution of errors the formula for calculation of allowable error for each input of the summator has the form: $\varepsilon_i = \varepsilon / (n \cdot \alpha_i)$, where $\varepsilon$ is the summator allowable error, $n$ is the number of inputs of the summator, $\alpha_i$ is a synapse weight of the corresponding input of the summator.

Under priority distribution of errors, at first the errors are attributed to those inputs which are most significant according to some criterion, and then the rest of the summator allowable error is distributed between the other inputs uniformly or proportionally.

Similarly it is possible to calculate allowable errors for input signals of the summator of any standard neuron if the errors for the result of work of the summator are known.

A special case of the adaptive summator considered above is the simple summator, in which the input signals are summarized without weights. For it the allowable error of each input is calculated using the formula: $\varepsilon_i \leq \varepsilon / n$.

For the adaptive summator it is possible to calculate both allowable errors of input signals of the summator and allowable errors of synapse weights. As well as for an allowable errors of input signals, for calculation of allowable errors of synapse weights it is possible to use uniform, proportional and priority distributions of errors. Under uniform distribution the allowable errors for synapse weights are calculated using the formula: $\varepsilon_i \leq \varepsilon / \sum_{i=1}^{k} |x_i|$, where $x_i$ is input signals of the summator.

Under proportional distribution the allowable errors for synapse weights are calculated using the formula: $\varepsilon_i = \varepsilon / (n \cdot x_i)$, where $n$ is the number of inputs of the summator, $x$ is the input signals of the summator.

Under priority distribution at first allowable errors are attributed to those inputs which are most significant according to some criterion, and then the remainder part of the summator allowable error is distributed between the other inputs uniformly or proportionally.

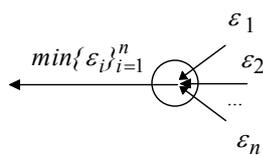

Fig. 3

Now we know how to calculate the error for any elements of standard neuron by the back-propagation of accuracy.

1) *Branch point*. If under the back-propagation of accuracy the incoming errors of the branch point are $\varepsilon_1, \varepsilon_2, ..., \varepsilon_n$ then as the error coming through the branch point one should choose $\min\{\varepsilon_i\}_{i=1}^{n}$ (Fig.3).

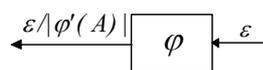

Fig.4

2) *Nonlinear transformer*. Let under direct functioning the input signal of the nonlinear transformer is equal to $A$, the output signal is equal to $y$, its allowable error is equal to $\delta$ and the nonlinear transformer has activation function $\varphi$. If under back-propagation the incoming error is $\varepsilon$, then after the nonlinear transformer we have the error $\varepsilon / \max|\varphi'(x)|$ where $x \in [\varphi^{-1}(y - \delta), \varphi^{-1}(y + \delta)]$ or in linear approximation $\varepsilon / \varphi'(A)$ (Fig.4).

3) *Adaptive summator*. If under back-propagation of accuracy to the output of the adaptive summator the incoming error is $\varepsilon$, then the error of each input of the summator should not exceed $\varepsilon_i$ (Fig.5).

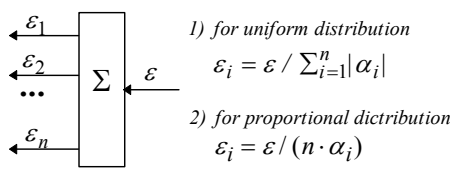

1) *for uniform distribution*
$$\varepsilon_i = \varepsilon / \sum_{i=1}^{n} |\alpha_i|$$

2) *for proportional dictribution*
$$\varepsilon_i = \varepsilon / (n \cdot \alpha_i)$$

*Fig.5*

When knowing how to calculate allowable errors for all elements of standard neuron, one can calculate allowable errors of signals for the whole network. If allowable errors for output signals of network are given, it is possible to calculate allowable errors for the last layer of the network. When the allowable errors of all input signals of the last layer of the network are calculated, we pass to calculation of allowable errors of previous layer and so on.

The back-propagation of accuracy can be applied not only to networks of layered structure, but also to cyclic and fully connected networks. Considering a step of functioning of a network as a layer, we "unfold" recurrent feed forward and fully connected networks to a network of layered structure. We calculate allowable errors for elements of standard neurons of each layer. Then we "roll up" the layered network to the initial. Since each layer of obtained networks is actually a step of functioning, for each element of a network at different steps we obtain different allowable errors. As allowable error for each element of the network the minimum of allowable errors over all steps is chosen.

We have considered how allowable errors of elements of networks are calculated for one example from training set. To calculate allowable errors of elements of network over all training set, it is necessary to calculate allowable errors for each example, and then for each element of the network as allowable error the minimal allowable error over all examples is chosen.

If we deal not with training set but with the range of values of input data, for calculation of allowable errors of elements over the whole range it is necessary to estimate the maximum of activation function of nonlinear transformers and for determination of allowable error of the input signal of nonlinear transformer to use in the formula the maximum of the activation function over the whole range: $|\max_\Omega \varphi|$ where $\Omega$ is the range of values of input data.

Thus, the solution of the problem of calculation of allowable errors for each element of network by back-propagation of accuracy is surprisingly similar to a back-propagation of error, but with other rules of passing through the elements. The method allows to formulate the requirements to the accuracy of calculations and to the realization of technical devices, if the requirements to the accuracy of output signals of the network are known.

**References**

1. Gorban A.N. *Traning Neural Networks*.- Moscow, USSR-USA JV "ParaGraph" 1990. – 160 pp. (Russian) (English Translation: AMSE Transaction, Scientific Siberian, A, 1993, Vol. 6. Neurocomputing. Tassin (France): AMSE Press. PP.1-134).
2. *Neuroinformatic and its applications* // Materials of 3 All-Russian Seminar, October 6-8, 1995 Part 1 / Krasnoyarsk State Technical University Press. Krasnoyarsk. – 1995. – 229 p.
3. Kimura T., Shima T. *Synapse weight accuracy of analog neuro chip* // Proceedings of International Joint Conference on Neural Networks. – Japan, Nagoya, October 25-29, 1993. – Vol.1. – P. 891-894.
4. Anguita D., Ridella S., Rovetta S. *Limiting the effects of weight errors in feed forward networks using interval arithmetic* // Proceedings of International Conference on Neural Networks (ICNN'96). – USA, Washington, June 3-6, 1996. – Vol.1. – P. 414-417.
5. Edwards P., Murray A. *Modelling weight- and input-noise in MLP learning* // Proceedings of International Conference on Neural Networks (ICNN'96). – USA, Washington, June 3-6, 1996. – Vol.1. – P. 78-83.